\documentclass[letterpaper]{article} 
\usepackage[]{aaai25}
\usepackage{times}  
\usepackage{helvet}  
\usepackage{courier}  
\usepackage[hyphens]{url}  
\usepackage{graphicx} 
\usepackage{natbib} 
\usepackage{booktabs}
\usepackage{caption} 
\usepackage{algorithm}
\usepackage{algorithmic}

\usepackage[skip=5pt]{caption}  

\usepackage{xcolor} 
\usepackage{tikz}
\usetikzlibrary{shapes.geometric, arrows.meta, positioning}

\tikzstyle{process} = [rectangle, draw=black, thick, minimum height=1.2cm, minimum width=2.8cm, text centered]
\tikzstyle{data} = [cylinder, shape border rotate=90, draw=black, minimum height=1.5cm, minimum width=1cm, aspect=0.25, thick]
\tikzstyle{arrow} = [thick, ->, >=Stealth]
\tikzstyle{textblock} = [align=left, font=\small]
\usepackage{newfloat}
\usepackage{listings}

\usepackage{caption}
\usepackage{adjustbox}

\newcommand{\dname}[1]{{TeleScope}}

\DeclareCaptionStyle{ruled}{labelfont=normalfont,labelsep=colon,strut=off} 
\floatstyle{ruled}
\newfloat{listing}{tb}{lst}{}
\floatname{listing}{Listing}
\usepackage{xcolor}

\newcommand{\answerYes}[1]{\textcolor{blue}{#1}} 
\newcommand{\answerNo}[1]{\textcolor{teal}{#1}} 
\newcommand{\answerNA}[1]{\textcolor{gray}{#1}}

\title{TeleScope: A Longitudinal Dataset for Investigating Online Discourse and Information Interaction on Telegram}
\author {
    Susmita Gangopadhyay\textsuperscript{\rm 1},
    Danilo Dess{\'i}\textsuperscript{\rm 2},
    Dimitar Dimitrov\textsuperscript{\rm 1},
    Stefan Dietze\textsuperscript{\rm 1,3} 
}
\affiliations {
    \textsuperscript{\rm 1} GESIS – Leibniz Institute for the Social Sciences, Cologne, Germany\\
    \textsuperscript{\rm 2} Department of Computer Science, College of Computing and Informatics, University of Sharjah, Sharjah, UAE \\
    \textsuperscript{\rm 3} Heinrich Heine University  Düsseldorf, Düsseldorf, Germany \\
    susmita.gangopadhyay@gesis.org, ddessi@sharjah.ac.ae, dimitar.dimitrov@gesis.org, stefan.dietze@gesis.org
}

\usepackage{bibentry}

\begin{document}

\maketitle

\begin{abstract}
Telegram is a globally popular instant messaging platform known for its strong emphasis on security, privacy, and unique social networking features. It has recently emerged as the host for various cross-domain analysis and research works, such as social media influence, propaganda studies, and extremism. 
This paper introduces \dname{}, an extensive dataset suite that, to our knowledge, is the largest of its kind. It comprises metadata for about $500K$ Telegram channels and downloaded message metadata for about $71K$ public channels, accounting for around $120M$ crawled messages. We also release channel connections and user interaction data built using Telegram's message-forwarding feature to study multiple use cases, such as information spread and message-forwarding patterns. In addition, we provide data enrichments, such as language detection, active message posting periods for each channel, and Telegram entities extracted from messages, that enable online discourse analysis beyond what is possible with the original data alone. 
The dataset is designed for diverse applications, independent of specific research objectives, and sufficiently versatile to facilitate the replication of social media studies comparable to those conducted on platforms like X (formerly Twitter).
\end{abstract}

\section{Introduction}
In recent years, Telegram has emerged as one of the most widely used messaging platforms worldwide, distinguished by its unique combination of messaging and social media features. The platform hosts a wide range of communities covering personal, social, and political interests. Telegram’s popularity has been bolstered by its emphasis on user privacy and minimal moderation. Although this has become appealing to privacy-conscious users, it also allows for controversial content ~\cite{yayla2017Telegram} and rhetoric~\cite{walther2021us}.

In addition to its messaging features, Telegram channels provide one-to-many communication, enabling a single administrator or a small group of people to broadcast messages, updates, and announcements to potentially vast audiences across the platform\footnote{Telegram FAQ: \url{https://telegram.org/faq#q-what-39s-the-difference-between-groups-and-channels}}.  This unique structure maintained in these channels allows content to reach a high number of users quickly, amplifying the spread of news or opinions across the platform. Furthermore, unlike traditional social media tools that emphasize bidirectional or group interactions, these channels facilitate the dissemination of information with minimal or no direct feedback from followers. Consequently, Telegram channels can influence public opinion due to the limited room for dialogue or critical discussion as these channels' structure permits only their owners to post messages, forcing a single-direction flow of information. Their ability to serve as hubs for information propagation makes them pivotal for studying societal narratives, particularly in understanding how information is shared and framed in several communities and how influential channels can bias opinions or even contribute to spreading misinformation at scale~\cite{schulze2022far,key}. Thus, Telegram has recently become a focal point for discussions around societal and political issues, making monitoring and analysis of its content essential.

However, the current structure of Telegram presents significant challenges in accessing its content, analyzing the spread of information, and tracking how content propagates across the platform. This limitation arises because Telegram does not provide detailed data on the source and destination channels of forwarded messages. Users can only observe the number of times a message has been forwarded from a single channel, with the forwarding count varying across different channels. Further, Telegram content is not indexed by search engines, making it relatively inaccessible and opaque to users who are not already part of the communities or channels that utilise it. This hinders research and understanding of the platform, for example, in terms of the dynamics of information flow, the formation of communities, and the mechanisms driving the spread of both accurate information and misinformation.  

In parallel, X (formerly Twitter) has historically been a primary platform for social science research, with many algorithms and frameworks specifically designed to process and analyze X data~\cite{mccormick2017using,qudar2020tweetbert}. These frameworks have enabled researchers from various disciplines to conduct a wide range of social media studies, from the spread of information~\cite{zaman2010predicting} to the dynamics of online communities~\cite{ovadia2009exploring}. However, applying these algorithms directly to Telegram data is not feasible due to the structural differences between the platforms. For instance, unlike X, Telegram does not have features like retweet counts or collective statistics for views and forwards across messages. These distinctions necessitate tailored approaches for analyzing Telegram data. Moreover, due to the recent restrictions on X's APIs, accessing its data for academic research has become increasingly challenging, prompting the need for alternative data sources in social science research. 
This has made platforms like Telegram, with its growing user base and unique features, a new source for researchers looking to explore online discourse. However, existing datasets~\cite{solopova2021Telegram,hohn2022belelect} typically focus on collecting data from specific channels for targeted research purposes or lack unified features related to forwards, reactions, and views across channels~\cite{baumgartner2020pushshift,la2023tgdataset}, limiting their flexibility in supporting long-term social media studies similar to those conducted on platforms like X.

To address the limitations of existing datasets, with this paper, we introduce \dname{}\footnote{\url{https://data.gesis.org/telescope/}}, which contains the largest publicly available, multilingual, longitudinal, and continuous collection
from Telegram, including data enrichments and  novel
user interaction and message propagation information on a
per-channel and per-message basis. To make the dataset both transparent and valuable to the research community, we have implemented the following measures:
(i) \textit{public seed list}, i.e., we openly share our seed list to establish the provenance of our data and enhance its credibility.
(ii) \textit{continuous and longitudinal data}, i.e., our dataset captures continuous, longitudinal data to allow in-depth analysis over time.
(iii) \textit{comprehensive data coverage}, i.e., we collect data from the channels without imposing any restrictions on the number of messages, ensuring a complete dataset.
(iv) \textit{data enrichments}, i.e., in addition to the raw data, we provide data enrichments that offer additional insights, making them especially useful for researchers.
(v) \textit{novel features for network analysis}, i.e., we include innovative features designed to facilitate the study of Telegram’s network dynamics and interactions. In summary, our contribution includes:
\begin{itemize}
    \item \textbf{Enriched Channel Metadata (cf. Section~\ref{sec:cnl})}:
    The largest collection of Telegram channel metadata 
    comprising $534,137$ channels discovered through a snowball sampling approach. This dataset is designed to serve as a registry for selecting channels tailored to specific research needs. Beyond the source data, we provide data enrichments, including language detection for public channels and hourly message activity trends. 
    \item \textbf{Enriched Crawl of Telegram Message Metadata (cf. Section 4.2)}: We release metadata for approximately $120$ million crawled messages 
    from  $71,048$ public channels within these $534,137$ channels. This metadata is enriched with detected Telegram entities such as bold, italics, URLs, links, and hashtags. 
    
    \item \textbf{User Interaction Data and Cross-Channel Message Propagation Flows (cf. Section 4.3) }: As a unique contribution of our corpus, we provide channel connections, message propagation, and user interaction data. 
    We track the origins of forwarded messages,  map the path traveled by each message and provide insights into how channels connect and interact within the larger Telegram network through a channel-to-channel interaction graph and how far a message has been forwarded by message forwarding flows. We also release separate aggregated user interaction data that includes statistics like views, forwards, and reactions for forwarded messages across the entire network. 
        
\end{itemize}
\vspace{-2mm}

\section{Background and Related Work}
Telegram, launched in 2013, has grown into a widely-used instant messaging platform, amassing over 800 million active users by 2021 \footnote{Telegram FAQ: \url{https://Telegram.org/faq\#q-what-is-Telegram-what-do-i-do-here}}. The platform enables users to share various types of content, including text messages, images, audio, stickers, videos, and files up to 2 GB in size. Beyond private one-on-one chats, Telegram offers two additional modes of communication: Groups and Channels. Groups are designed for many-to-many interactions, whereas channels facilitate one-to-many communication. 
In channels, only administrators can post messages, while subscribers can only react and leave comments on posted messages. Channels are similar to groups and can be either public or private. However, there are notable distinctions, such as they can have an unlimited number of subscribers; however, subscriber information, such as nicknames, is not displayed, and only the total subscriber count is visible. These features have made Telegram channels a preferred tool for broadcasting content to large audiences. Public figures and organizations frequently use official channels to share updates and news. 

The Telegram ecosystem has been the subject of extensive research, with many studies leveraging Telegram data to address various cross-domain research problems. Recent works have demonstrated the diversity of research applications using Telegram data. For instance, Kloo et al.~\cite{kloo2024cross} collected data from both X and Telegram to analyze and compare discussions surrounding a specific campaign on these platforms. Similarly, Hoseini et al.~\cite{hoseini2024characterizing} focused on Telegram data to study information propagation within fringe communities, exploring how public groups and channels exchange messages across the Telegram network. Telegram data has been widely used to study various topics, such as tracking information flow across different languages in media outlets~\cite{hanley2024partial}, analyzing scientific visibility~\cite{venancio2024unraveling}, and understanding user coordination and political mobilization during elections or major political events in specific countries~\cite{venancio2024unraveling}. Some researchers have also published Telegram datasets for broader use. For instance, TGDataset, introduced by~\cite{la2023tgdataset}, contains data from one hundred twenty thousand channels. Similarly, authors in~\cite{baumgartner2020pushshift} created the Pushshift Telegram dataset, which includes about $27K$ channels and $317$ million messages from about $2.2M$ unique users. Telegram data has also been published for specialized studies, such as the \textit{BelElect} dataset for bias research on ``dark'' platforms~\cite{article1},  misinformation detection~\cite{sosa-sharoff-2022-multimodal}, and studying hate speech, offensive language, and online harm~\cite{solopova2021Telegram}.
 Although many studies use Telegram data, there is a lack of data availability at a large scale and over extended periods, making it hard for researchers to expand research findings. Even with available datasets, they are usually incomplete and tailored to specific purposes.
To overcome these limitations, we introduce our dataset suit \dname{}, the most extensive publicly available collection of Telegram data to date. It is multilingual, longitudinal, continuous, and includes data enrichments and detailed insights into user interactions and message propagation captured at both channel and message levels.

\section{Data Collection Methodology}
\begin{figure}[t]
\centering
\includegraphics[width=0.85\linewidth]{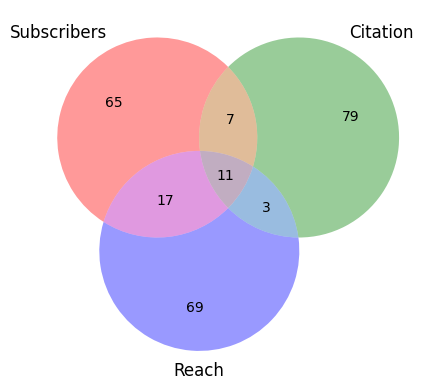}
\caption{Overlap of discovered channels across different seed selection criteria.}
\label{seed_list}
\end{figure}

This section presents our seed list creation for the channels (cf. Section~\ref{seedlist}), the data collection methodology (cf. Section~\ref{datacollection}), and an estimation of the coverage of the downloaded data (cf. Section~\ref{discoveryandcoverage}).

\subsection{Seed List Channel Collection}\label{seedlist}
Telegram does not provide a comprehensive 
and central directory of all channels.  
Therefore, we have built a seed list 
gathering channel IDs from a publicly available registry that organizes and ranks 
channels based on their metrics. 
We have utilized TGStat\footnote{https://tgstat.com/}, a popular free premium service that indexes over $150K$ Telegram channels and gathers statistics about them. TGStat is one of the most widely used registry of channels used in recent research related to Telegram~\cite{la2023tgdataset,alvisi2024unraveling,tikhomirova2021community}.
The channel selection for our seed list is based on three distinct criteria to ensure diverse coverage:

\begin{itemize}
    \item Number of subscribers \textit{i.e.}, the number of users who have opted to follow a Telegram channel. Channels with a large number of subscribers are included to ensure the analysis covers influential spaces with broad audiences.

    \item Number of citations \textit{i.e.},  the number of mentions or references made by other channels or users. These channels are chosen as they are more likely to play a central role in disseminating information across the platform.

    \item Reach \textit{i.e.}, the number of unique users who have viewed a message from the channel. This captures the extent of a channel's exposure and influence within the Telegram ecosystem.

\end{itemize}

We have collected the top $100$ channels from TGStat for each criterion (subscribers, citation, and reach), and compiled a unique list of $251$ channels in various languages, accounting for overlaps. Note that some channels appear in the top $100$ based on multiple criteria, resulting in fewer than $300$ unique channels. An overview of each collected channel set and their overlapping channels is presented in Figure~\ref{seed_list}.

\subsection{Data Collection Process and Statistics}\label{datacollection}

\begin{figure*}[t]
\centering
\includegraphics[width=0.8\linewidth]{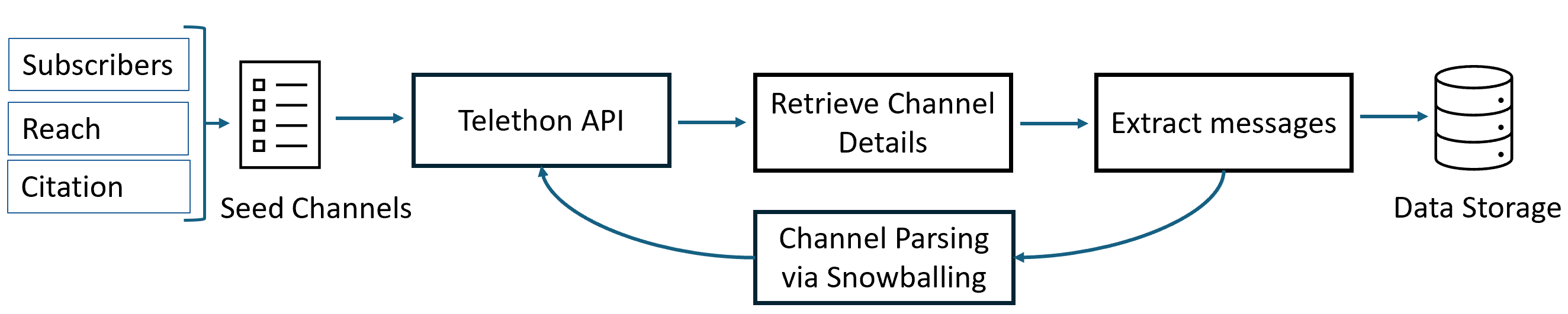}
\caption{ Depicts the data collection pipeline. Channels collected using three criteria are merged into a seedlist, enriched via the Telethon API, expanded through snowballing, and stored in a database.}
\label{Data_Collection}
\end{figure*}

Our data collection on Telegram is channel-centric, involving the extraction of metadata from all channels and message content from public channels only. We collect both content and metadata from channels 
using Telethon\footnote{https://docs.telethon.dev/en/stable/}, a Python interface for
the Telegram API. The Telethon API provides metadata for both channels and messages. Our approach leverages Telegram's ``message forwarding'' functionality. More precisely, we look at the source channel of forwarded messages within channels to discover new channels that can be included in our channel list. This process is iterative, thus enabling continuous expansion of the channel list. 
The collection is initiated with the seed list of $251$ unique channels. Whenever we encounter a message forwarded from a channel not already in our list, we add that channel, collect its data, and recursively follow all channels it interacts with. 
For channels, it includes data such as title, creation date, unique identification number, and various channel settings (e.g., usage configurations,
administrator restrictions, and whether the channel is a bot). For each public channel, we download all messages till Oct 2024. We do not collect any images and videos to avoid copyright issues or sensitive content. Messages can either be original content posted to a channel or can be forwarded content from another channel or user. Each message has several accompanying metadata, such as a unique identification number within a channel, the date/time the message is sent, whether it includes media (e.g., images or video), whether the user is a bot, whether they are a verified user, etc. Through the Telethon API, we collect all metadata available from the API without any particular selection. We stored data and metadata for each channel in a PostgreSQL relational database management system and the messages in a MongoDB database.  Our complete data collection pipeline is illustrated in Figure~\ref{Data_Collection}. 
The data collection for \dname{} ran from February 2024; the first static snapshot was taken in October 2024. The statistics of the first static snapshot of the data is represented in Table~\ref{tab:statistics}. At the time of taking the snapshot, our crawlers were actively discovering new channels; these will be leveraged for further releases of \dname{}. By this stage, we had identified and indexed over $1.2$ million channels, with the metadata collection for these channels and their associated messages still ongoing.

\begin{table}[t]
\centering
\small 
\renewcommand{\arraystretch}{1.0} 
\setlength{\tabcolsep}{0pt}       

\begin{tabular}{l c}
\toprule
\textbf{Feature} & \textbf{Value} \\
\midrule
Time frame & Feb 1, 2024 – Oct 29, 2024 \\
Discovered channels & $1{,}210{,}272$ \\
Channels with downloaded metadata & $534{,}137$ \\
Fully downloaded public channels & $71{,}048$ \\
Number of downloaded messages & $120{,}024{,}020$ \\
Avg. messages per channel & $1{,}689.33$ \\
Percentage of forwarded messages & $19.6$\% \\
Avg. messages downloaded/hour & $20{,}495$ \\
Complete dataset size & $76$ GB (zipped) \\
\bottomrule
\end{tabular}
\caption{Telegram data collection statistics.}
\label{tab:statistics}
\end{table}

\begin{figure}[t]
\centering
\includegraphics[width=0.8\linewidth]{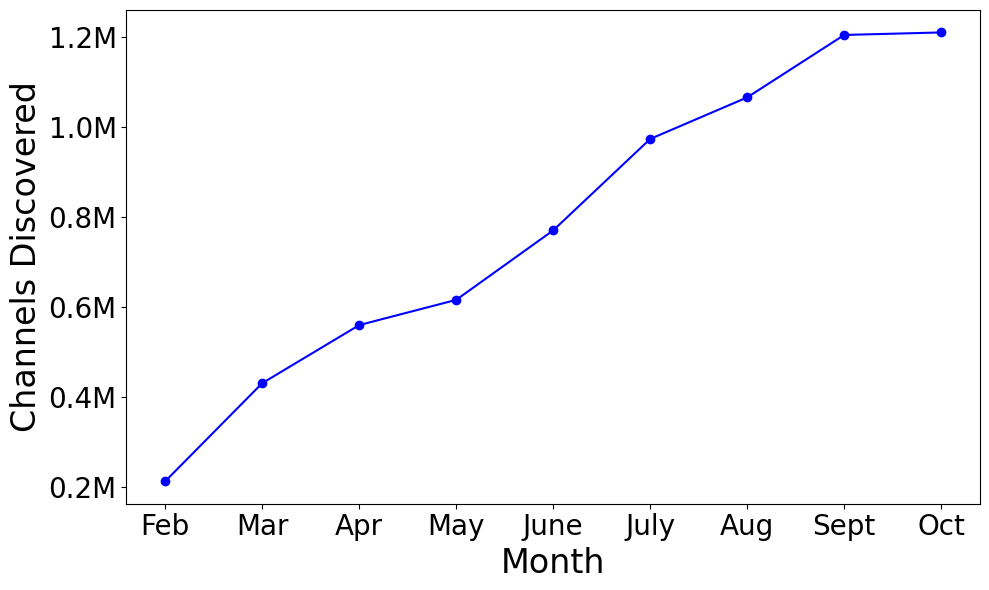}
\caption{Channels discovered via snowball sampling over time.}
\label{snowball}
\end{figure}

\subsection{Channel Discovery and Data Coverage}\label{discoveryandcoverage}
As mentioned above, Telegram does not provide a central directory for tracking all channels, therefore, estimating the coverage of the downloaded data is essential. Figure~\ref{snowball} illustrates the monthly discovery of channels using the snowball sampling approach. Our crawler identified approximately $200$K new channels each month, with a decline in the final month, suggesting a decreasing expansion in discovery. 
To better understand the coverage of our dataset, we also monitored the public channels discovered and downloaded through each criterion used to compose our seed list. As observed in Figure~\ref{seed_list}, the overlap between the channels selected on different criteria is initially minimal. However, as the process progressed, a significant overlap emerged among the channels discovered by each. Out of the $71$k channels that we downloaded, the initial set of channels selected using the subscriber criterion discovered $64,889$ channels, reach uncovered $64,817$, and citation found $64,803$, with $64,791$ channels overlapping across all three. This shows that the crawling process ultimately converges to discover similar channel sets regardless of the criterion used. This can occur because certain channels act as hubs, leading to a convergence in the channels discovered throughout the rest of the graph. This finding suggests that while different criteria may provide distinct channel lists as seeds for channel discovery, their long-term impact on dataset expansion diminishes as the process progresses.

\section{Dataset Description} 
This section describes \dname{} dataset structure, organized in enriched channel metadata (cf. Section~\ref{sec:cnl}), enriched message metadata (cf. Section~\ref{sec:msg}), and cross-channel message propagation flows (cf. Section~\ref{sec:cid}). 
\subsection{Enriched Channel Metadata}\label{sec:cnl}
This section describes the metadata collected for Telegram channels, including key attributes and data enrichments such as channel language and temporal metadata.
\vspace{-1mm}
\subsubsection{Channel Metadata.} The Channel Metadata includes metadata for all the channels fully downloaded in our dataset ($534, 137$) organized into a single JSON file.  The dataset contains information about each channel and includes details such as username, creation date, number of subscribers, whether a channel is fake or scam, whether there are restrictions imposed on the channel, etc. The details for the fields included in the channel metadata file are available in the official documentation\footnote{ \url{https://tl.telethon.dev/types/chat full.html}}.
\vspace{-2mm}
\subsubsection{Channel Language.}
We compute the primary language of Telegram channels using the Python Langdetect library\footnote{https://pypi.org/project/langdetect/}, a language detection tool that supports the identification of 55 languages and stores in a CSV file. According to the library’s standard documentation\footnote{\url{https://www.slideshare.net/slideshow/language-detection-library-for-java/6014274}}, it can detect the language of texts with over 99\% precision. The library employs a Naive Bayes algorithm and character n-grams as features to classify texts into language categories. It is open-source and has been tested on over $9,000$ news articles with an accuracy of $99.77$\%. To ensure meaningful results when analyzing text-based messages, we extract all messages from a channel, combine them into a long text, and then apply language detection. If the text contains more than one language, the library attempts to identify the language that is most prevalent or best matches the majority of the text. In cases where the language is not recognized by the library or there is no predominant language, the output shows ``cannot determine". In our dataset, out of $71,048$ channels, there are $59$ channels, the language of which could not be determined.
\vspace{-4mm}
\subsubsection{Temporal Channel Metadata.}
Temporal Channel Metadata is a CSV file that includes information about the distribution of messages across daily hours within a single channel, allowing for an analysis of the most active time intervals for specific channels. Additionally, it includes the creation date of each channel, offering insights into when channels first appeared on the platform. The lifespan of a channel is also recorded, defined as the time span between its creation and its most recent recorded activity.

\subsection{Enriched Crawl of Telegram Message Metadata}\label{sec:msg}
This section presents the metadata collected for Telegram messages and data enrichments such as telegram entities. 
\vspace{-2mm}
\subsubsection{Message Metadata.} It contains $71,048$ newline delimited JSON files, each corresponding to a channel, containing metadata of all its messages. Each channel is stored as a zipped folder containing all message details as provided by the Telethon API\footnote{\url{https://tl.telethon.dev/constructors/message.html}}, including forwards, reactions, views, the date and time the message was sent and replies. 
\vspace{-2mm}
\subsubsection{Telegram Entities.} Telegram provides entities that enable specific platform functionalities. According to official Telegram documentation\footnote{https://docs.telethon.dev/en/stable/concepts/entities.html}, an entity refers to any user, chat, or channel object that the Telethon API may return in response to certain methods. They are structural components within a message that add semantic meaning, formatting, or interactivity to its content. Examples include mentions (\textit{e.g}., @username) that refer to specific users, hashtags (\textit{e.g}., topics) to label discussions, and URLs that link to external resources. Additionally, entities include text formatting elements such as bold, italic, and underline for structured presentation. These entities not only enhance the expressiveness of messages but also enable researchers to focus only on the most informative parts of a message.  Instead of explicitly providing the entity, Telegram shares only the span or location of the entity within the text. To enhance usability, we enrich the data by extracting the entities for each message, converting raw spans into ready-to-use data. By releasing the extracted entities, our dataset provides a means to access and analyze the content without the need to have access to the full message text. \vspace{-2mm}

\subsection{User Interaction Data and Cross-Channel Message
Propagation Flows}\label{sec:cid}
This section describes how messages propagate between channels and the resulting interaction dynamics in the form of a graph. We also provide aggregated user interactions with messages to support comprehensive information spreading and user behavior analyses.

\subsubsection{Message Forwarding Flows.}
\begin{figure*}[t]
    \centering
  \includegraphics[width=1.05\linewidth]{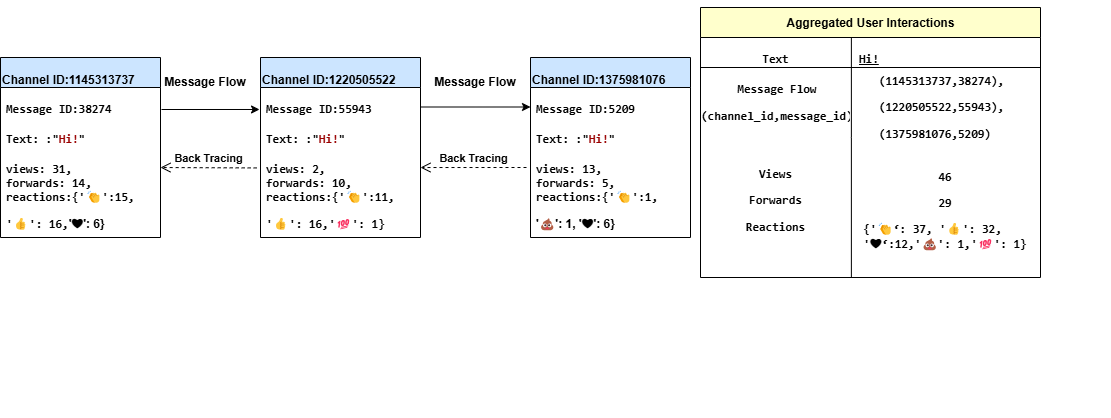}
  \vspace{-60pt}
    \caption{An example of message forwarding flows in \dname{}.}
   \label{fig:forwarding_flow_2}
       \vspace{-1.5em}
\end{figure*}

Telegram does not provide a built-in method for tracking the full forwarding history of messages beyond their immediate source. Our dataset addresses this limitation by including Message Forwarding Flows. Figure~\ref{fig:forwarding_flow_2} shows an example of a message forwarding flow,  capturing the entire path a message takes as it is forwarded from one channel to another, recording both the source and destination of the forward. For instance, in Figure~\ref{fig:forwarding_flow_2}, the message ID \textit{38274} from channel \textit{1145313737} propagates in the Telegram network, becoming the message with ID \textit{5209} in channel \textit{1375981076} in the end. 
When a message $m^{c_A}_{i}$ from channel $c_A$ (the source) is forwarded and becomes message $m^{c_B}_{j}$ in channel $c_B$ (the destination), only $m^{c_B}_{j}$ retains the origin information. As a result, the original message $m^{c_A}_{i}$ from channel $c_A$ lacks direct information about its connection to other channels. To address this, we leverage the wideness of our dataset to build this missing link by looking at the message $m^{c_B}_{j}$ in the destination channel $c_B$ and backward tracing the source message and channel creating a path from channel $c_A$ to $c_B$ in, representing the source and destination respectively in the flow. 

It is important to note that the source message and its channel may not always be present in the input data (\textit{i.e.}, its content might not be part of the current dataset), which indicates that expanding the dataset may be necessary for a more comprehensive analysis. Message propagation statistics are shown in Table~\ref{tab:message_propagation}. For all messages within our dataset, flows are constructed using this method of backward-tracing the forwarding chains. The length of the forwarding chain does not reflect the global number of forwards within the Telegram platform, but it reflects the length of the path a message traveled within our dataset.  

\begin{table}[t!]
\centering

\begin{tabular}{@{}lc@{}}
\toprule

Total number of messages                    & $31,227,109$                   \\
Number of unique messages                    & $308,147$                   \\
Smallest message flow                    & $2$                    \\
Longest message flow                    & $4,810$                   \\
Average message flow                    & $2.54$                   \\
\bottomrule
\end{tabular}
\caption{Message propagation statistics in \dname{}.}
\label{tab:message_propagation}
\end{table}
\subsubsection{Channel-to-Channel Graph.} 
\begin{figure}[t]
    \centering
    \includegraphics[width=0.7\linewidth]{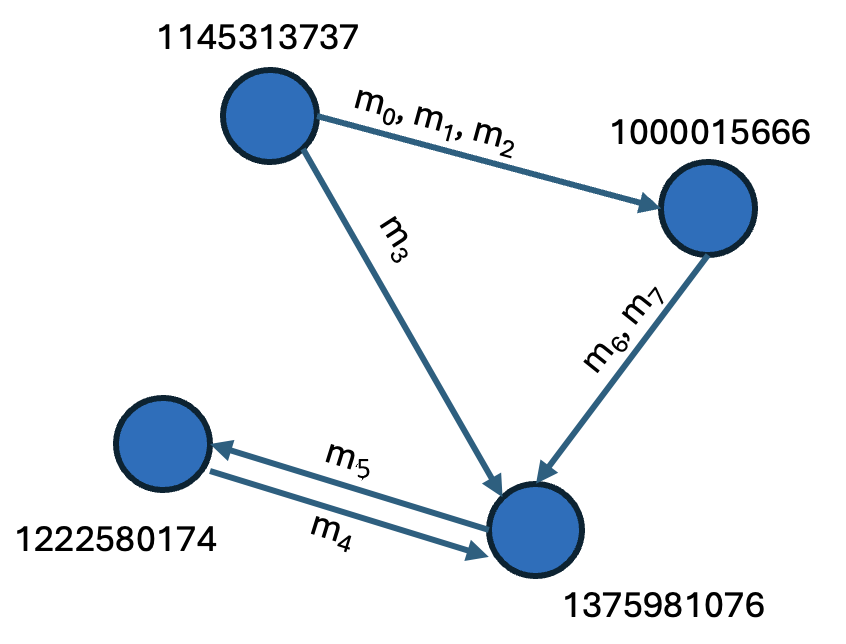}
    \caption{Example of channel-to-channel graph.}
    \label{fig:Channel To Channel Graph}
\end{figure}

Research into online communication often requires understanding how information spreads across different communities. 
The channel-to-channel graph in our dataset shows how Telegram channels are connected through message forwarding. In this graph, channels are shown as nodes, and each forwarding event is represented by a directed edge. 
It tracks all messages sent from one channel (source) to another (destination) and adds the message ID from the source channel to each edge. 
Figure~\ref{fig:Channel To Channel Graph}  illustrates an example of this graph, where each node represents a channel labeled by its ID, and edges capture the forwarding of messages between channels. If a message is forwarded from channel $c_A$ to channel $c_B$, the graph includes all such messages (e.g., $m_{0}$, $m_{1}$, \dots, $m_{i}$) along the path $c_A \rightarrow c_B$. In Figure~\ref{fig:Channel To Channel Graph} messages $m_0, m_1, m_2$ are forwarded from channel \textit{1145313737} to channel \textit{1000015666}, message $m_5$ from channel \textit{1375981076} to channel \textit{1222580174}, message $m_4$ from channel \textit{1222580174} to channel \textit{1375981076}, etc. The dataset combines the source channel, destination channel, and message IDs of the source channel to record all messages shared between them. It is a valuable tool for researchers aiming to study how information spreads across multiple channels, track how specific channels gain momentum, and identify key channels that influence the flow of information.  
The dataset contains $261,171$ channel nodes and $2,733,720$ edges. Notably, this graph has been constructed using forwarding data from the fully downloaded $71K$ dataset. Consequently, it may also include channels that have been discovered through forwarding relationships but have not yet been downloaded. 

\subsubsection{Aggregated User Interactions.}
Users who are not channel admins cannot directly post messages in Telegram channels, but they can still engage with the content through actions such as viewing, forwarding, reacting, etc. The Telegram API does not natively offer aggregated statistics for metrics such as views, forwards, and reactions across its entire network but only provides those at the channel level (cf. Figure~\ref{fig:forwarding_flow_2}).  Since user interactions per message are crucial for many research questions, \textit{e.g.}, by providing indicators of virality, popularity, or perceived authority of specific claims, \dname{} addresses this shortcoming by computing and consolidating these metrics for a specific message throughout the network of channels present in the dataset. 
For instance, consider again the message with ID \textit{38274} from channel \textit{1145313737}, which becomes the message
with ID \textit{5209} in channel \textit{1375981076} while taking two hops in our dataset. For these two hops, the message received a number of views, forwards, and reactions in a total of three channels. By summing up these numbers, we compute aggregated scores for these metrics at the dataset level.
Although the aggregated scores do not provide Telegram-wide statistics, this aggregation enables a more detailed analysis of a message's reach and engagement, offering insights into its performance within Telegram's ecosystem at the \dname{} dataset level. By capturing this comprehensive data, users can better understand the extent of the message's influence and interaction levels across the network. 

\section{Data Access and FAIR Principles}
We publicly released the \dname{} dataset adhering to FAIR principles~\cite{wilkinson2016fair}. Our dataset is \textbf{Findable} and available on Archiving BASIS\footnote{\url{https://www.re3data.org/repository/r3d100011062}}, an open repository managed by  GESIS Data Services with a DOI:\url{https://doi.org/10.7802/2825}. The dataset is \textbf{Accessible} and made of three components: channel metadata and enrichments, message metadata with enrichments, and user interaction and channel propagation data. Users can easily dive into the parts of the data that matter most to them. All the information is \textbf{Interoperable} and encoded in JSON and CSV formats, ensuring compatibility with established data representation and exchange standards. We also released Telegram crawler information and scripts that can replicate the enrichments on a public GitHub repository \footnote{\url{https://github.com/susmita3107/TeleScope}}, making our work \textbf{Reusable} for the users.

To ensure long-term sustainability, GESIS, as a research data infrastructure organization, commits to hosting the data and providing updates over time, as is done for X~\cite{fafalios2018tweetskb} and online claims~\cite{gangopadhyay2024investigating}.  Access to sensitive raw message data can be requested in a secure and privacy-aware manner through the GESIS' Secure Data Center\footnote{\url{https://www.gesis.org/en/services/processing-and-analyzing-data/analysis-of-sensitive-data/secure-data-center-sdc}}.

\section{Dataset Analysis} This section presents key statistics from \dname{} to give deeper insight into the collected data and its possible applications.

\subsection{Language Distribution}
We examine the language coverage of the dataset to gain insights into the identified channels. Table~\ref{tab:language_percentages} summarizes the distribution of languages among the $71,048$ public channels for which we downloaded the messages. The prevalence of Russian suggests that the dataset primarily captures content from Russian-speaking communities, though the presence of multiple languages highlights its diverse nature. While Russian is the most predominant language, the actual linguistic landscape of Telegram as a whole may be more diverse than what is captured here, as the language distribution may differ for public channels that have not been discovered yet with our snowball sampling approach and for private channels, which could not be collected. 

\begin{figure}[t!]
\centering
\includegraphics[width=0.85\linewidth]{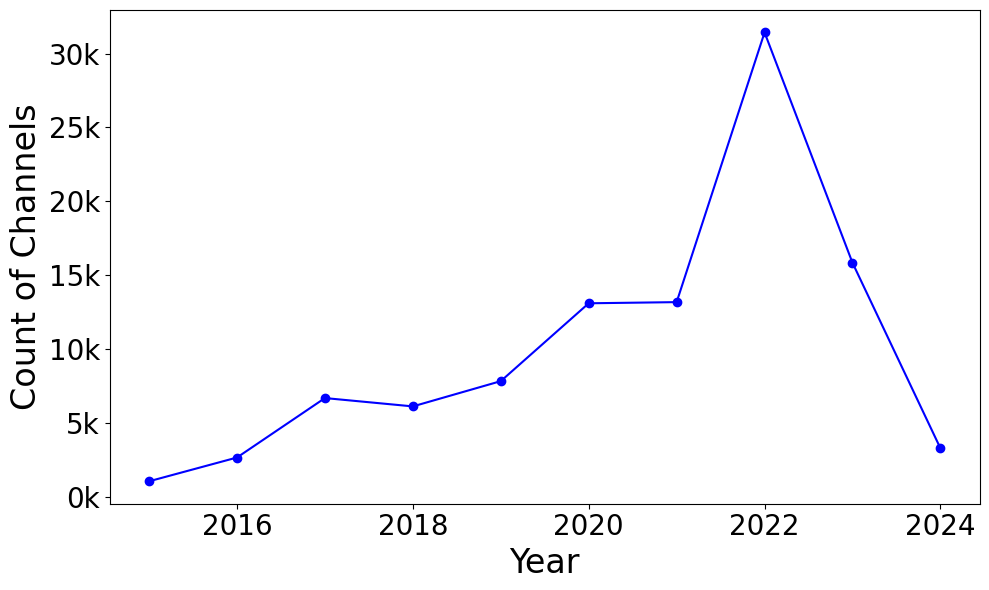}
\caption{Channels created per year within \dname{} data.}
\label{fig:channelCreation}
\end{figure}

\subsection{Channel Analysis}
From the metadata collected for about $500K$ channels, \dname{} dataset encompasses channels created as early as $2015$ up to $2024$ (cf. Figure~\ref{fig:channelCreation}). A significant portion of the channels have been established after $2021$. This trend may be linked to the migration of users from WhatsApp to Telegram following concerns over WhatsApp’s updated privacy policy~\footnote{\url{https://www.whatsapp.com/legal/privacy-policy}}. Such findings highlight the potential for social scientists and researchers to leverage this dataset to explore shifts in user behavior, communication patterns, and the broader dynamics of social media ecosystems. Insights into channel creation trends allow researchers to analyze the growth of Telegram’s channel-based communication ecosystem over time and its correlation with events, such as Telegram policy changes or global socio-political events. By examining channel metadata such as creation dates, names, and entities, researchers can identify patterns in channel diversity and specialization, uncovering how specific types of content or communities emerge and evolve. 

\begin{table}[t!]
\centering

\begin{tabular}{@{}lc@{}}
\toprule
\textbf{Language}     & \textbf{Percentage (\%)} \\ \midrule
ru                    & 82.29                   \\
uk                    & 4.6                    \\
en                    & 4.2                    \\
fa                    & 2.2                    \\
de                    & 1.1                    \\
cannot determine      & 0.08                    \\
others                & 5.53                    \\ \bottomrule
\end{tabular}
\caption{Language distribution among downloaded public channels in \dname{}.}

\label{tab:language_percentages}
\vspace{-2mm}
\end{table}

In our analysis, we also examined specific properties associated with the channels fetched through the Telethon API, such as whether the channels are marked as fake, scam, verified, or restricted (cf. Table~\ref{tab:property_percentages}). A channel receives a ``verified'' status when it can prove its authenticity across at least two major social media platforms\footnote{https://Telegram.org/verify} (\textit{e.g}., TikTok, Facebook, Twitter, Instagram). On the other hand, a ``scam'' status is assigned to channels and groups that have been reported for fraudulent activities by multiple users\footnote{Jack Ricle. Aug. 2022. Scammers in Telegram and how to report. https://www.
Telegramadviser.com/scammers-in-Telegram-and-how-to-report/}. ``Fake" channels are those that impersonate well-known people or services~\cite{inproceedings} and for each ``restricted'' channel the Telethon API provides a reason, \textit{e.g}. ``adult" or ``copyrighted" content.  Although our dataset includes a relatively small number of channels flagged as fake, scam, restricted or verified channels, these properties still provide valuable insights for studying fraudulent or deceptive channels within Telegram. Additionally, by identifying restricted channels, researchers can explore the reasons behind their restrictions to obtain a deeper understanding of how Telegram manages potentially harmful content. This analysis opens up opportunities for future research into understanding content moderation on the platform.

\vspace{-2mm}
\begin{table}[t!]
\centering
\begin{tabular}{lcc}
\toprule
\textbf{Property} & \textbf{True\% } & \textbf{False\%} \\
\midrule
Fake        & 0.01\%  & 99.99\% \\
Scam        & 0.03\%  & 99.97\% \\
Verified    & 0.52\%  & 99.48\% \\
Restricted  & 0.14\%  & 99.86\% \\
\bottomrule
\end{tabular}
\caption{Percentage distribution of selected binary properties in the \dname{} channels.}
\label{tab:property_percentages}
\end{table}


\begin{figure}[htbp]
\centering
    \includegraphics[width=0.8\linewidth]{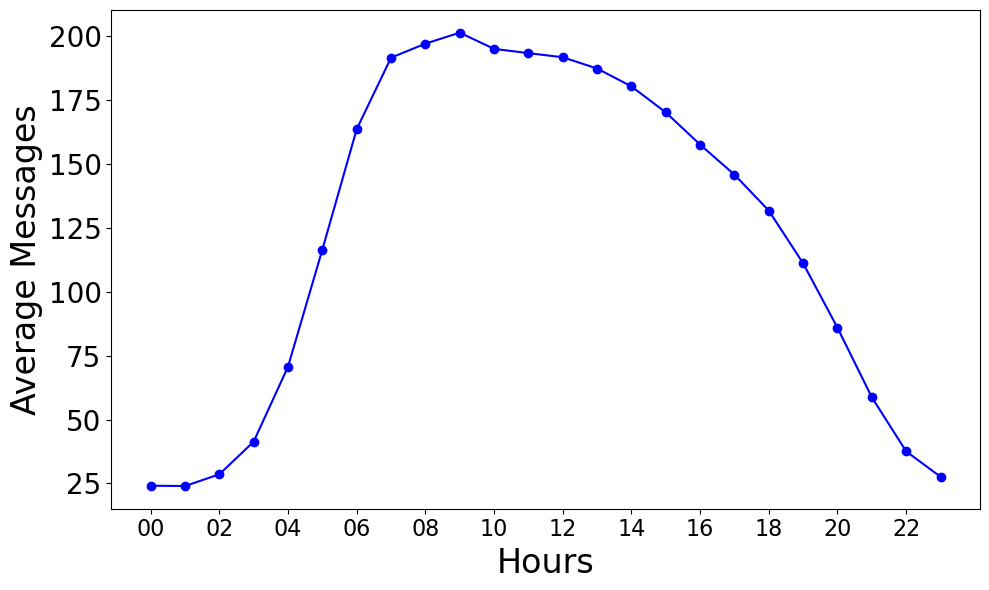}
    \caption{ Active periods reporting the average number of messages per hour. Time Zones are presented in UTC.}
    \label{Busy_hours}
\end{figure}

\subsection{Dominant Active Periods}

We studied the message posting pattern of the public channels present in our dataset. Figure~\ref{Busy_hours} illustrates the average number of messages sent per hour throughout a day within these $71K$ channels. According to our analysis, we notice a distinct pattern in messaging activity, with the lowest activity occurring during the early morning hours (Hour 00 to Hour 04). The activity begins to rise sharply from Hour 05, peaking between Hour 06 and Hour 09. The highest activity is during Hour 09. After reaching this peak, the average number of messages remains relatively stable during the mid-morning and early afternoon hours before gradually declining from Hour 16 onward. By the late evening (Hour 22 to Hour 23), activity levels return to their lowest point. This pattern could suggest that Telegram users are most active during morning hours, potentially reflecting work-related or routine activities, and least active during late-night hours. \vspace{-2mm}

\subsection{Hashtags}

\begin{figure}[t!]
\centering
   \includegraphics[width=0.8\linewidth]{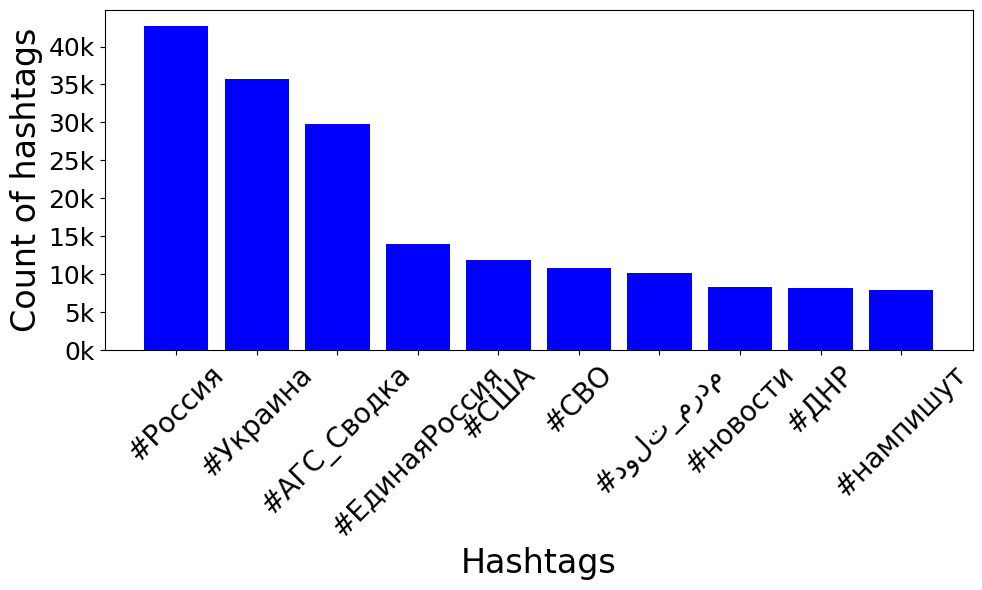}
    \caption{ The top 10 hashtags in the messages.}
    \label{hashtags}
\end{figure}

Figure~\ref{hashtags} depicts the top 10 hashtags used in the dataset, along with their corresponding counts. The most frequently used hashtag translated in English from the Figure is Russia, which significantly surpasses others with over $40K$ occurrences. This is followed by Ukraine and Artillery Fire Summary, each with over $30K$ mentions, indicating their prominence in the dataset. The next set of hashtags, including United Russia and USA, shows a steady decline in usage, with counts ranging between $15K$ and $20K$. The remaining hashtags, such as news and Donetsk People's Republic, display relatively lower frequencies, averaging around $10K$ mentions each. Notably, the presence of hashtags in different languages, including Arabic, highlights the multilingual nature of the dataset. The dominance of certain hashtags suggests a strong focus on geopolitical and regional topics, particularly related to Russia, Ukraine, and associated themes.

\subsection{User Interactions}

\begin{figure*}
\centering
   \begin{tabular}{ccc}
     \includegraphics[width=6cm,height=4cm]{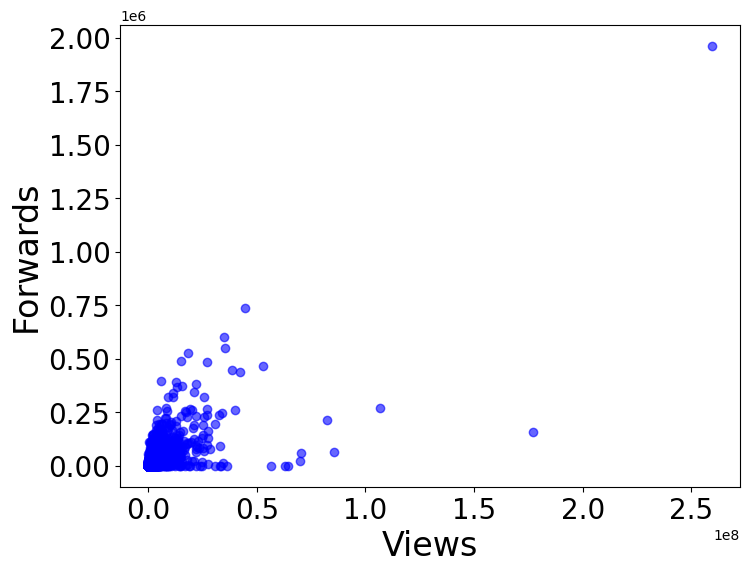}    &  
        \includegraphics[width=6cm,height=4cm]{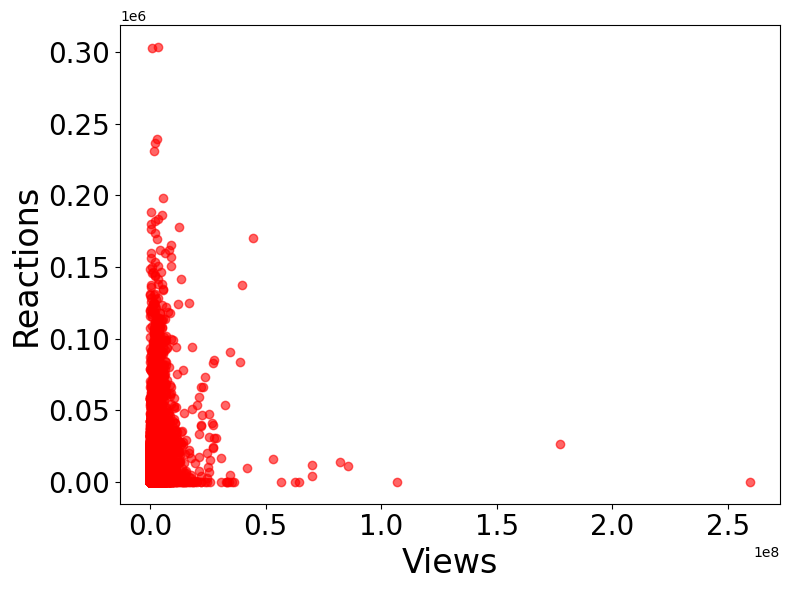} 
   \end{tabular}   
\caption{ The correlation between 
  (a) views and forwards on the left, (b)   views and reactions on the right.}
\vspace{-2mm}
\label{fig:correlation}
\end{figure*}

We also show the relation between some descriptive metadata regarding metrics of the downloaded messages such as views and forwards in Figure~\ref{fig:correlation}(a), and views and reactions in Figure~\ref{fig:correlation}(b). 
In the scatter plots, the majority of data points are clustered at the lower ends of the axes, indicating that most content has low views while also having low forwards and reactions. Although there are a few outliers with significantly higher values, these are exceptions rather than the norm. The plots suggest that higher views do not consistently translate to higher forwards or reactions, highlighting a relatively weak correlation between views and engagement metrics. This demonstrates variations in how audiences engage with content.

\section{\dname{} Use Cases}
This section provides an overview of \dname{} and its potential use cases. The dataset ensemble is created to maintain a multilingual, continuous longitudinal data collection process, free from specific focus or bias in its collection. This approach ensures that the dataset remains versatile and broadly applicable, allowing researchers to replicate studies across overlapping domains such as social science, natural language processing (NLP), network science, etc. 

\subsection{ Replication of Social Media Research}
The availability of aggregated data in Telegram, such as the number of forwards and reactions (\textit{e.g.}, likes), provides a unique opportunity to replicate and adapt research methodologies that have been developed for other social media like Twitter. For example, on Twitter, retweets have been a cornerstone for studying various topics, including information diffusion~\cite{10.1145/3610058} (\textit{e.g}., understanding how content spreads through social networks), virality prediction~\cite{10.1145/3543873.3587373} (\textit{e.g}., identifying factors that make content widely shared), community structure analysis~\cite{ellis2024social} (\textit{e.g}., detecting clusters and influencers within social networks), bias detection and analysis of particular entities~\cite{xiao2023detecting}. Similarly, the forwarding feature in Telegram serves as a proxy for information propagation, enabling the application of these frameworks to explore how messages flow across channels and communities. This can be studied either using the channel-to-channel graph for studies focusing on the Telegram network, or exploring the message flows for studies focusing on the spread of specific types of messages. Furthermore, the inclusion of reaction data in Telegram expands the scope of analysis, allowing researchers to investigate topics such as user engagement, sentiment analysis, and content popularity. By adapting these existing methodologies to Telegram's ecosystem, it becomes possible to analyze pressing topics such as misinformation spread and the dynamics of public discourse on a platform that emphasizes channel-based communication. \vspace{-2mm}
\subsection{Network and Community Discovery}
Our derived datasets offer a range of valuable use cases. The channel-to-channel graph allows researchers to track the spread of information across Telegram, making it a powerful tool for studying misinformation and fake news. By analyzing this graph, researchers can detect communities, identify key channels that act as information hubs, and uncover patterns in message dissemination. Visualizing the graph helps social scientists gain insights into how information flows, making it useful for tasks like media monitoring, influence detection, and understanding the dynamics of communication. Researchers can explore topics transferred between channels, discover clusters of interconnected channels, and analyze changes in channel interactions over time, such as why one channel most often forwards from certain channels, or why certain channels have stopped communicating with others. The Message Forwarding Flows can further enhance these analyses by mapping the complete propagation path of messages. This enables detailed studies on the temporal dynamics of message forwarding, and trends in content virality, providing a deeper understanding of how information spreads and whether specific types of messages are spread more within the Telegram network. \vspace{-2mm}

\subsection{Entity-Based Search and Exploration}
The entities in the \dname{} dataset, comprising Telegram-specific entities such as hashtags, mentions, and URLs, enrich the dataset and unlock diverse use cases. Entities enable data to be searched and discovered within Telegram channels. By leveraging these entities, researchers can categorize channels based on their thematic focus, aligning with specific research objectives. This enables tailored investigations, such as analyzing political discourse by tracking mentions of political figures or parties, tracking trends in specific regions through references to specific locations, or studying the impact of particular hashtags 
and shared URLs on content dissemination and engagement. External links, combined with entities, facilitate research on information flows between Telegram and the web, including the analysis of misinformation propagation or campaigns.

\subsection{Data Source for Low-Resource Language}
Our dataset reveals a multilingual landscape, covering 47 different languages, with Russian (ru), Ukrainian (uk), English (en), Iranian (fa), and German (de) emerging as the most prominent languages. In addition to these widely spoken languages, we also include channels in several low-resource languages such as Tagalog (tl), Swahili (sw), Catalan (ca), Punjabi (pa), Somali (so), and Lithuanian (lt), among others. These low-resource languages, while culturally significant, are often underrepresented in computational linguistics, lacking the extensive tools, research, and training data available in comparison to more widely spoken languages like English or Chinese. Channels in these languages can be particularly difficult to find and study elsewhere, which makes our dataset a valuable resource for NLP research in these regions. Furthermore, these channels offer unique opportunities to track communication patterns and information flow within lesser-known linguistic communities, providing researchers with a rich source of data for developing language technologies and studying social interactions in these underrepresented languages. 
\vspace{-2.5mm}
\section{Conclusion and Future Work} \vspace{-1mm}
Telegram has become highly popular in recent years, making it essential to study and understand the activities on the platform. This work presents \dname{}, a dataset suit featuring metadata for around $500K$ Telegram channels and message metadata from around $71K$ public channels within these $500K$ channels. 
To build \dname{}, we started with $251$ seed channels and used a snowball sampling method to expand the dataset. This approach links channels through message forwarding, allowing us to uncover over $1.2M$ channels, with the potential to discover more in future iterations. Alongside the metadata provided by Telegram, \dname{} includes enriched metadata and novel information on channel interactions, offering a more comprehensive view of the platform's dynamics.
\dname{} represents a snapshot of Telegram channels, capturing informative metadata to perform social science analysis and research. 
To address privacy concerns, we release only message metadata, excluding the actual message content. 
Looking ahead, we aim to continue data collection, expand metadata coverage, and release updated versions of the dataset regularly. Specifically, we plan to publish a new dataset snapshot every year, integrating updated statistics and metadata. We are also currently in the process of discovering multilingual named entities (\textit{people, organizations, locations, dates}) and mentions within messages. 
This will provide researchers with up-to-date data for future analyses, enabling the study of temporal changes in channel characteristics and the evolution of Telegram communities over time. \dname{} dataset is intended to accommodate a wide variety of research needs across different domains, along with the ability to replicate  prior works on platforms such as X.
\vspace{-3mm}

\section*{Acknowledgments} We thank Massimo La Morgia for fruitful discussions that contributed to shaping the initial idea behind this work.

 \vspace{-2mm}

\bibliography{aaai25}

\section*{Paper Checklist}

\begin{enumerate}

\item For most authors...
\begin{enumerate}
    \item  Would answering this research question advance science without violating social contracts, such as violating privacy norms, perpetuating unfair profiling, exacerbating the socio-economic divide, or implying disrespect to societies or cultures?
    \answerYes{Yes}
  \item Do your main claims in the abstract and introduction accurately reflect the paper's contributions and scope?
    \answerYes{Yes}
   \item Do you clarify how the proposed methodological approach is appropriate for the claims made? 
    \answerYes{Yes. Refer to Data Collection Methodology section}
   \item Do you clarify what are possible artifacts in the data used, given population-specific distributions?
    \answerYes{Yes}
  \item Did you describe the limitations of your work?
    \answerYes{Yes}
  \item Did you discuss any potential negative societal impacts of your work?
    \answerNA{NA}.
      \item Did you discuss any potential misuse of your work?
    \answerYes{Yes. In the conclusion section}
    \item Did you describe steps taken to prevent or mitigate potential negative outcomes of the research, such as data and model documentation, data anonymization, responsible release, access control, and the reproducibility of findings?
    \answerYes{Yes, in the Data Access section}
  \item Have you read the ethics review guidelines and ensured that your paper conforms to them?
    \answerYes{Yes}
\end{enumerate}

\item Additionally, if your study involves hypotheses testing...
\begin{enumerate}
  \item Did you clearly state the assumptions underlying all theoretical results?
    \answerNA{NA}
  \item Have you provided justifications for all theoretical results?
    \answerNA{NA}
  \item Did you discuss competing hypotheses or theories that might challenge or complement your theoretical results?
    \answerNA{NA}
  \item Have you considered alternative mechanisms or explanations that might account for the same outcomes observed in your study?
    \answerNA{NA}
  \item Did you address potential biases or limitations in your theoretical framework?
    \answerNA{NA}
  \item Have you related your theoretical results to the existing literature in social science?
    \answerNA{NA}
  \item Did you discuss the implications of your theoretical results for policy, practice, or further research in the social science domain?
    \answerNA{NA}
\end{enumerate}

\item Additionally, if you are including theoretical proofs...
\begin{enumerate}
  \item Did you state the full set of assumptions of all theoretical results?
   \answerNA{NA}
	\item Did you include complete proofs of all theoretical results?
    \answerNA{NA}
\end{enumerate}

\item Additionally, if you ran machine learning experiments...
\begin{enumerate}
  \item Did you include the code, data, and instructions needed to reproduce the main experimental results (either in the supplemental material or as a URL)?
    \answerNA{NA}
  \item Did you specify all the training details (e.g., data splits, hyperparameters, how they were chosen)?
    \answerNA{NA}
     \item Did you report error bars (e.g., with respect to the random seed after running experiments multiple times)?
   \answerNA{NA}
	\item Did you include the total amount of compute and the type of resources used (e.g., type of GPUs, internal cluster, or cloud provider)?
    \answerNA{NA}
     \item Do you justify how the proposed evaluation is sufficient and appropriate to the claims made? 
   \answerNA{NA}
     \item Do you discuss what is ``the cost`` of misclassification and fault (in)tolerance?
    \answerNA{NA}
  
\end{enumerate}

\item Additionally, if you are using existing assets (e.g., code, data, models) or curating/releasing new assets, \textbf{without compromising anonymity}...
\begin{enumerate}
  \item If your work uses existing assets, did you cite the creators?
    \answerYes{Yes. We used the Telethon API and the paper is referenced.}
  \item Did you mention the license of the assets?
    \answerNA{NA}
  \item Did you include any new assets in the supplemental material or as a URL?
   \answerNA{NA}
  \item Did you discuss whether and how consent was obtained from people whose data you're using/curating?
    \answerNA{NA. We only collect public data.}
  \item Did you discuss whether the data you are using/curating contains personally identifiable information or offensive content?
    \answerYes{Yes. For this, we removed all raw data and we only release metadata. Please refer to Data Access section}
\item If you are curating or releasing new datasets, did you discuss how you intend to make your datasets FAIR (see \citet{fair})?
\answerYes{Yes. We provide a DOI and use established technologies to make the data FAIR. }
\item If you are curating or releasing new datasets, did you create a Datasheet for the Dataset (see \citet{gebru2021datasheets})? 
\answerNo{No. We will do so if the paper is accepted. }
\end{enumerate}

\item Additionally, if you used crowdsourcing or conducted research with human subjects, \textbf{without compromising anonymity}...
\begin{enumerate}
  \item Did you include the full text of instructions given to participants and screenshots?
     \answerNA{NA}
  \item Did you describe any potential participant risks, with mentions of Institutional Review Board (IRB) approvals?
      \answerNA{NA}
  \item Did you include the estimated hourly wage paid to participants and the total amount spent on participant compensation?
     \answerNA{NA}
   \item Did you discuss how data is stored, shared, and deidentified?
    \answerNA{NA}
\end{enumerate}

\end{enumerate}

\end{document}